\documentclass[%
reprint,
amsmath,amssymb,
aps,
]{revtex4-1}

\usepackage{graphicx}
\usepackage{dcolumn}
\usepackage{bm}
\usepackage{color}

\usepackage{url}

\begin{document}
	
\preprint{APS/123-QED}
	
\title{Comparative study of Hermitian and non-Hermitian topological dielectric photonic crystals}
	
\author{Menglin L. N. Chen}
\email{\url{http://www.zjuisee.zju.edu.cn/weisha/Publications/Files/FD_PCs_Wilson_loop.zip}}%
\affiliation{%
Department of Electrical and Electronic Engineering, The University of Hong Kong, Hong Kong
}
\author{Li Jun Jiang}
\email{jianglj@hku.hk}
\affiliation{%
Department of Electrical and Electronic Engineering, The University of Hong Kong, Hong Kong
}%
\author{Ran Zhao}
\affiliation{%
Key Laboratory of Intelligent Computing and Signal Processing, Ministry of Education, Anhui University, Hefei 230039, China 
}%
\author{Zhihao Lan}
\affiliation{%
Department of Electronic and Electrical Engineering, University College London, United Kingdom 
}%
\author{Wei E. I. Sha}
\email{weisha@zju.edu.cn}	
\affiliation{%
Key Laboratory of Micro-nano Electronic Devices and Smart Systems of Zhejiang Province, College of Information Science and Electronic Engineering, Zhejiang University, Hangzhou 310027, China
}%
\author{Shuang Zhang}
\affiliation{%
Department of Electrical and Electronic Engineering, The University of Hong Kong, Hong Kong
}%


\begin{abstract}
	
The effects of gain and loss on the band structures of a bulk topological dielectric photonic crystal (PC) with $C_{6v}$ symmetry and the PC-air-PC interface are studied based on first-principle calculation. To illustrate the importance of parity-time (PT) symmetry, three systems are considered, namely the PT-symmetric, PT-asymmetric, and lossy systems. We find that the system with gain and loss distributed in a PT symmetric manner exhibits a phase transition from a PT exact phase to a PT broken phase as the strength of the gain and loss increases, while for the PT-asymmetric and lossy systems, no such phase transition occurs. Furthermore, based on the Wilson loop calculation, the topology of the PT-symmetric system in the PT exact phase is demonstrated to keep unchanged as the Hermitian system. At last, different kinds of edge states in Hermitian systems under the influences of gain and loss are studied and we find that while the eigenfrequencies of nontrivial edge states become complex conjugate pairs, they keep real for the trivial defect states.

\end{abstract}

\maketitle

\section{Introduction}

Recently, synthetic non-Hermitian optical systems based on parity-time (PT) symmetry have attracted considerable attentions due to the unique optical effects they provide, which hold great promise for practical applications not achievable in conventional Hermitian optics \cite{pt_review}. The degeneracies in parameter space of non-Hermitian optical systems, i.e., spectral singularities known as exceptional points \cite{ep1,ep2,2014PRX,Fuliang2014topological}, possess unconventional features compared to their Hermitian counterparts, where not only the eigenvalues but also the corresponding eigenvectors of the underlying system coalesce simultaneously. Optical response around exceptional points shows counter-intuitive phenomena, such as, unidirectional invisibility \cite{Unidirectional_Invisibility_11PRL, Regensburger_12Nature}, single-mode lasing \cite{lasing_science14, orbit_lasing_16science}, enhanced sensitivity \cite{sensing_17nature, sensing2_17nature}, and enhanced spontaneous emission \cite{spontaneous_emission_16PRL}. For example, dynamically encircling an EP can lead to chiral behaviors, while encircling the EP in different directions results in different output states \cite{encirclingEP_16nature}. Also, a recent study \cite{ep_starting_18prx} shows that whether or not the dynamics is chiral actually depends on the starting point of the loop. Exceptional points have also been studied in high order or higher dimensions, for example, two-dimensional (2D) exceptional surfaces \cite{Exceptional_surfaces_optica19} or 3D exceptional hypersurface \cite{Exceptional_hypersurfaces_oe20}. Up to now, EPs have been studied in diverse systems, such as photonic crystals (PCs) \cite{ep_pc_15nature}, plasmonic waveguides \cite{ep_plasmonic_11oe}, and optomechanical systems \cite{optomech_16nature}.

On the other hand, non-Hermiticity can change the band topology in topology systems or even induce a topological phase transition from a trivial system to a topological counterpart \cite{Takata18PRL}. For example, non-Hermitian perturbations can deform a Dirac cone and spawn a ring of exceptional points rather than opening up a gap in Hermitian systems \cite{ep_pc_15nature}. Bulk Fermi arc and polarization half charge from paired exceptional points have also been observed \cite{half_18science}. A nontrivial Berry phase of the eigenstates can be obtained by encircling exceptional points \cite{Geometric_phase_12pra}. Furthermore, edge states around interface between non-Hermitian topological crystals show new features \cite{pt_interface_nm_17} and the effect of gain and loss on quantum-spin-Hall (QSH) edge states has attracted great interest recently~\cite{fan2019PT,zhang2021topological}. It is interesting to see how gain and loss can affect the trivial and nontrivial edge states in a dramatically different way and most importantly, to observe their difference (trivial and nontrivial) in a single system.

In this work, gain and loss are introduced into a dielectric PC with $C_{6v}$ symmetry that has been proved to mimic the QSH system and show nontrivial topology~\cite{huxiao2015scheme}. We analyze their band structures and topological invariants based on first-principle calculation. Three systems, namely the PT-symmetric, PT-asymmetric, and lossy systems, have been investigated. We provide a comprehensive study on their band structures and topological invariants for both non-Hermitian and Hermitian systems.  Moreover, we study the evolution of the trivial and nontrivial edge states under the influence of gain and loss in a single system.

\section{Bulk Band of non-Hermitian PCs}

\begin{figure}[htbp]
	\centering
	\includegraphics[width=\columnwidth]{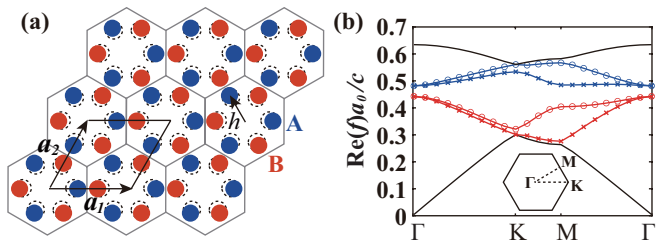}
	\caption{(a) Geometry of the 2D PC arranged in a deformed honeycomb lattice with six cylinders in each unit cell, where $\bm{a_1}$ and $\bm{a_2}$ are the translation vectors and $a_0$ is the lattice constant. Dielectric constants of the cylinders $\epsilon_A=12+ \gamma_A i$ and $\epsilon_B=12+ \gamma_B i$, where $\gamma_{A/B}$ is the strength of the loss or gain in A or B, depending on its sign. Other parameters are: $h=6$~mm,  $a_0=2.8h$ and the diameter of the cylinders is $4$~mm. (b) The band structure of the TM modes when $\epsilon_A=\epsilon_B=12$.}
	\label{cyl}
\end{figure}

A 2D PC arranged in a deformed honeycomb lattice with six cylinders in each unit cell is shown in Fig.~\ref{cyl}(a), where the original positions of the cylinders in a honeycomb lattice are indicated by the dashed circles. $\bm{a_1}$ and $\bm{a_2}$ are the two translation vectors and $a_0$ is the lattice constant. It has been shown that expansion of the six cylinders with respect to the center of the hexagonal unit cell results in a nontrivial topology of the PC~\cite{huxiao2015scheme}. In this paper, our aim is to understand how the introduction of gain and loss can lead to nontrivial phase transitions in the topological PC structure. We consider two types of cylinders, A and B, with the same radii but different dielectric constants. Figure~\ref{cyl}(a) shows one type of distributions of A and B.

We focus on the transverse-magnetic (TM) modes. The master equation is written as
\begin{equation}
	\nabla \times \nabla \times \mathbf{E}_z(\mathbf{r}) - k_0^2\epsilon(\mathbf{r})\mathbf{E}_z(\mathbf{r})= 0,
	\label{master}
\end{equation}
where $k_0$ is the free-space wave number and $\epsilon(\mathbf{r})$ is the position-dependent dielectric constant. Due to the periodicity of the PC, we can restrict the problem to a single unit cell [the black rhombus in Fig.~\ref{cyl}(a)] and as such the field component $E_z(\mathbf{r})$ could be expressed as Bloch states,
\begin{equation}
	E_\mathbf{k}(\mathbf{r}) = e^{i \mathbf{k} \cdot \mathbf{r}} u_\mathbf{k}(\mathbf{r}).
	\label{bloch}
\end{equation}
Here, $u_\mathbf{k}(\mathbf{r})$ are periodic functions with the same periodicity of the lattice, i.e., $u_\mathbf{k}(\mathbf{r})=u_\mathbf{k}(\mathbf{r}+\mathbf{R})$ for any lattice vector $\mathbf{R}$. By applying the generalized finite-difference (FD) method, Eq.~(\ref{master}) could be rewritten in a matrix form, $ME_z =k_0^2E_z$. The eigenequation can then be solved numerically to get the eigenmodes $E_z$ and eigenvalues $k_0^2$~\cite{menglinPRA,zhao2020first}. It should be noted that $M$ is not Hermitian when lossy and gain materials are included. The eigenmodes we defined and calculated are the right eigenmodes, if not mentioned otherwise.

\begin{figure}[htbp]
	\centering
	\includegraphics[width=\columnwidth]{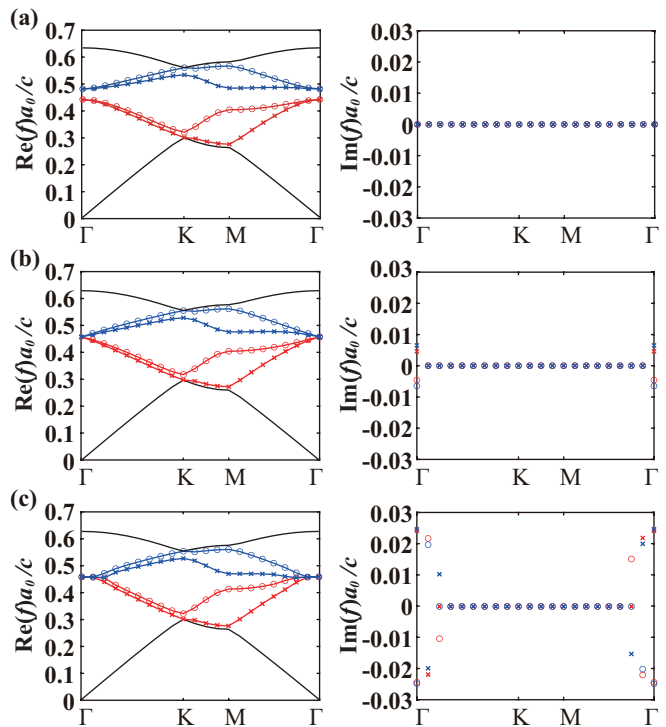}
	\caption{The complex band structure of the PT-symmetric PC system when (a) $\gamma_A=-\gamma_B=0.1$, (b) $\gamma_A=-\gamma_B=1.3$, and (c) $\gamma_A=-\gamma_B=2$.}
	\label{band_3case}
\end{figure}

When the PC structure is Hermitian ($\gamma_A=\gamma_B=0$), the eigenfrequencies of all the bulk bands are real. There is a complete band gap between two doubly degenerate bands at the $\Gamma$ point [Fig.~\ref{cyl}(b)]. When we add gain and loss with the same strength into A and B which are distributed as in Fig.~\ref{cyl}(a), the PC system preserves the PT symmetry. In Fig.~\ref{band_3case}, the complex band structures for three typical cases with different $\gamma$ are plotted. With a small $\gamma$, as shown in Fig.~\ref{band_3case}(a), the band gap remains and the frequencies are still real. The system is in the PT exact phase. At a specific value of $\gamma$, the bands at the lower and upper edges of the band gap begin to merge. The four modes coalesce into a single mode at the $\Gamma$ point, forming an exceptional point. As $\gamma$ increases further, the system enters the PT broken phase: the bulk gap closes as the frequencies of the upper and lower bands become ``zipping" complex conjugate pairs near the $\Gamma$ point. It should be noticed that from a tight-binding estimation, the critical $\gamma$ for the PT phase transition is equal to the difference of the inter- and intrasite hoppings, which is roughly $1.3$ in our case.

\begin{figure}[htbp]
	\centering
	\includegraphics[width=\columnwidth]{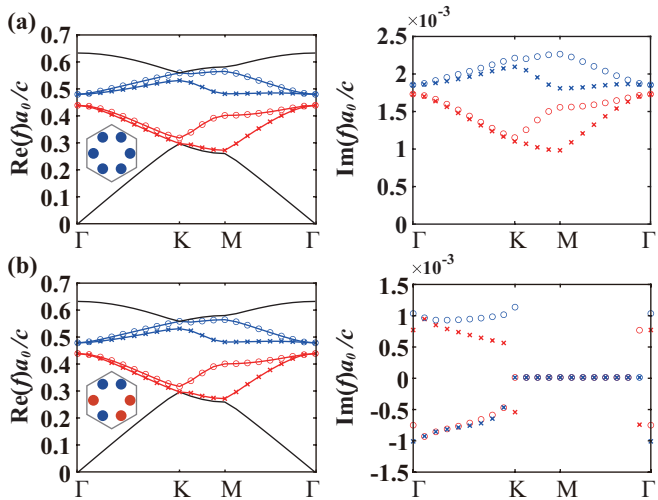}
	\caption{The complex band structure of the (a) lossy PC when $\gamma_A=\gamma_B=0.1$ and (b) PT-asymmetric PC when $\gamma_A=-\gamma_B=0.1$.}
	\label{loss}
\end{figure}

When the dielectric PC is purely lossy, i.e. $\gamma_A=\gamma_B>0$, we can see from its band structures in Fig.~\ref{loss}(a) that the imaginary parts of the frequencies are all positive, indicating that the eigenstates are all lossy. It is known that scaled dielectric function $\epsilon(\mathbf{r})*s$ results in scaled mode frequencies $\omega/\sqrt s$. Therefore, the added loss introduces a complex scaling factor so that the imaginary part of the band structure presents the same trend as in the real part. By distributing the cylinders in a different manner, the PT-asymmetric PC is analyzed as shown in Fig.~\ref{loss}(b). Different from the PT-symmetric PC, the eigenfrequencies become complex as soon as the gain and loss are introduced. Their imaginary parts have positive and negative values but they are not conjugate pairs.

\begin{figure}[htbp]
	\centering
	\includegraphics[width=\columnwidth]{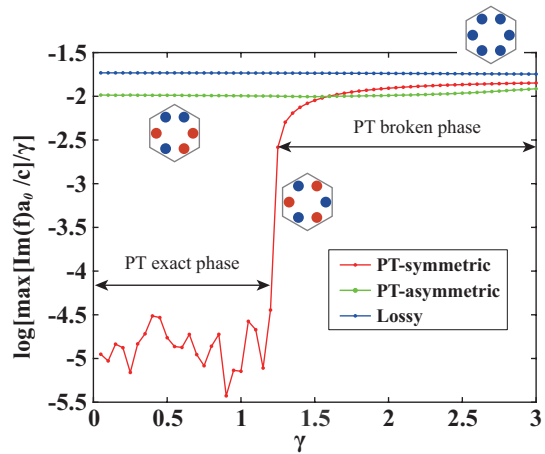}
	\caption{The evolution of the imaginary part of the band structure as a function of $\gamma$ for the PT-symmetric ($\gamma=\gamma_A=-\gamma_B$),  PT-asymmetric, and lossy $(\gamma=\gamma_A=\gamma_B)$ systems.}
	\label{pt_phase}
\end{figure}

In Fig.~\ref{pt_phase}, we present the evolution of the imaginary parts of the frequencies for the PT-symmetric, PT-asymmetric, and lossy systems as $\gamma$ increases, where the maximum imaginary part of the four doubly degenerate bands divided by $\gamma$ is shown. From the results, we can see that the PT-symmetric system exhibits a PT phase transition with a threshold at around $\gamma=1.3$, while the PT-asymmetric and lossy systems show a thresholdless transition and continuous evolution as $\gamma$ increases. This phenomenon is consistent with the PT-symmetric quantum spin Hall system~\cite{fan2019PT}.

\section{Wilson loop characterization}

\begin{figure}[htbp]
	\centering
	\includegraphics[width=\columnwidth]{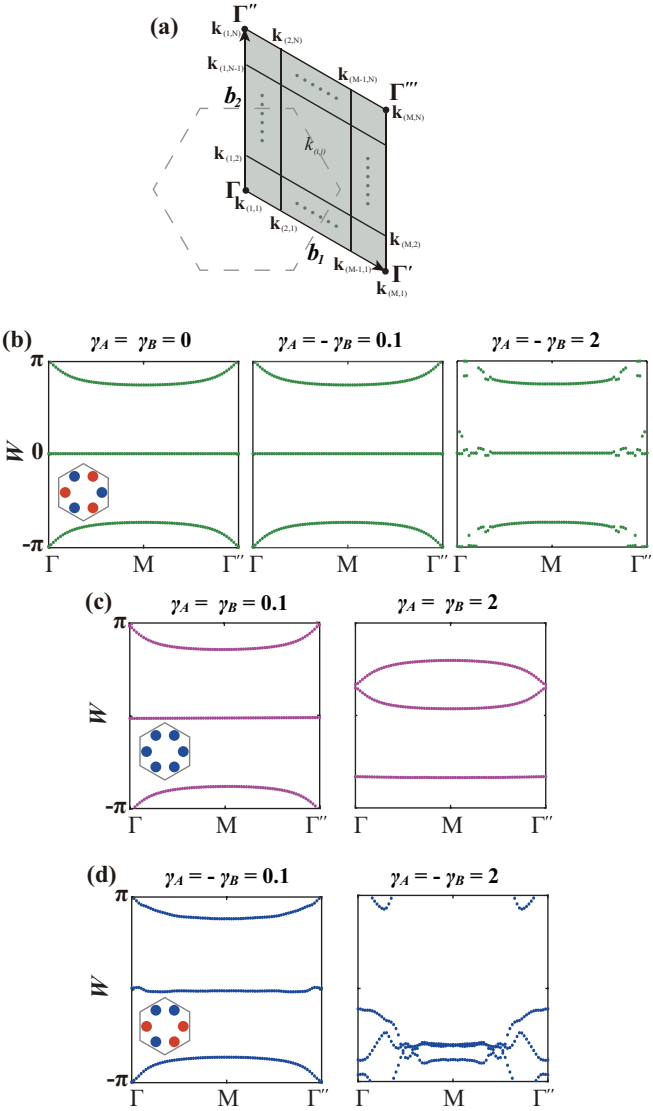}
	\caption{(a) The discretized BZ and Wilson loops for the (b) PT-symmetric, (c) lossy, and (d) PT-asymmetric PC systems.}
	\label{WL}
\end{figure}

Chern number is used as the topological invariant of non-Hermitian systems with broken time symmetry~\cite{silv2020first}. For topological systems with spin degree of freedom, the Wilson loop classifies their different topological properties~\cite{WL_Jiang}. To calculate the Wilson loop, first, we discretize the first Brillouin zone (BZ) along the reciprocal lattice vectors, $\bm{b_1}$ and $\bm{b_2}$, as shown in Fig.~\ref{WL}(a). The Wilson loop can be calculated at each discretized $\mathbf{k}$ point, swept from $\Gamma$ to $\Gamma'$ or $\Gamma$ to $\Gamma''$.

For isolated bands, the Wilson loop along $\bm{b_2}$ is obtained using
\begin{equation}
	W(\mathbf{k}_{(1,j)})= \operatorname{Im} \left[ \ln \prod_{m=1}^{M-1}
	U_{\mathbf{k}_{(m,j)} \rightarrow \mathbf{k}_{(m+1,j)}}
	\right], \\
	j=1,2,...,N.
	\label{calWL}
\end{equation}
Here, $U_{\mathbf{k}_{\alpha} \rightarrow \mathbf{k}_{\beta}} = \frac {\left\langle u_{\mathbf{k}_{\alpha}} | u_{\mathbf{k}_{\beta}}\right\rangle} {\left|\left\langle u_{\mathbf{k}_{\alpha}} | u_{\mathbf{k}_{\beta}} \right\rangle\right|}$, and $\left\langle  u_{\mathbf{k}_{\alpha}} | u_{\mathbf{k}_{\beta}} \right\rangle = \int \epsilon(\mathbf{r}) u_{\mathbf{k}_{\alpha}}^{*} (\mathbf{r}) \cdot u_{\mathbf{k}_{\beta}}(\mathbf{r}) d^2 \mathbf{r}$. $u_{\mathbf{k}}(\mathbf{r})$ is calculated by using our FD solver.

The Wilson loop is gauge invariant. Therefore, in numerical calculation, although an arbitrary phase is added at each calculated eigenstate, it will not alter the results. For example, if we discretize the BZ into four plaquettes, the multiplication in \eqref{calWL} can be written as $\left\langle U_1^R|U_2^R\right\rangle \left\langle U_2^R|U_3^R \right\rangle \left\langle U_3^R|U_4^R \right\rangle \left\langle U_4^R|U_5^R \right\rangle$. Here $R$ means we are calculating the right eigenvectors. Apparently, the states within the BZ show up twice during the multiplication, so their phase factors cancel. Similarly, we can use left eigenvectors, i.e., $\left\langle U_1^L|U_2^L\right\rangle \left\langle U_2^L|U_3^L \right\rangle \left\langle U_3^L|U_4^L \right\rangle \left\langle U_4^L|U_5^L \right\rangle$. Alternatively, combination of right and left eigenvectors can also be adopted. Since the phases for right and left eigenvectors are different, for the states at the same locations, the same eigenvectors should be chosen. For example, by using $\left\langle U_1^R|U_2^L\right\rangle \left\langle U_2^L|U_3^R \right\rangle \left\langle U_3^R|U_4^L \right\rangle \left\langle U_4^L|U_5^R \right\rangle$ and $\left\langle U_1^L|U_2^R\right\rangle \left\langle U_2^R|U_3^L \right\rangle \left\langle U_3^L|U_4^R \right\rangle \left\langle U_4^R|U_5^L \right\rangle$, correct results can also be obtained. These findings are consistent with the claims in~\cite{Fuliang2014topological}. Besides of the phases of the states within the BZ, the phases at the two boundaries should also be taken into consideration. The states at the boundaries are equivalent, i.e. $E_{\mathbf{k}_{(1,j)}}(\mathbf{r}) = E_{\mathbf{k}_{(M,j)}}(\mathbf{r})$. From \eqref{bloch}, we have
\begin{align}
\begin{split}
E_{\mathbf{k}_{(1,j)}}(\mathbf{r}) 
&= 
e^{i \mathbf{k}_{(1,j)} \cdot \mathbf{r}} u_{\mathbf{k}_{(1,j)}}(\mathbf{r}),\\
E_{\mathbf{k}_{(M,j)}}(\mathbf{r}) 
&=
e^{i \mathbf{k}_{(M,j)} \cdot \mathbf{r}} u_{\mathbf{k}_{(M,j)}}(\mathbf{r}).
\label{EK}
\end{split}
\end{align}
Therefore,
\begin{align}
\begin{split}
u_{\mathbf{k}_{(M,j)}}(\mathbf{r})
&=
E_{\mathbf{k}_{(M,j)}}(\mathbf{r})e^{- i \mathbf{k}_{(M,j)} \cdot \mathbf{r}}\\
&=
u_{\mathbf{k}_{(1,j)}}(\mathbf{r})e^{i (\mathbf{k}_{(1,j)} - \mathbf{k}_{(M,j)}) \cdot \mathbf{r}}\\
&=
u_{\mathbf{k}_{(1,j)}}(\mathbf{r})e^{-i \bf{b}_2 \cdot \mathbf{r}},
\label{uK}
\end{split}
\end{align}
which means the states at the boundaries should vary by a fixed phase factor in spite of the arbitrary phases added from the numerical solver. However, when \eqref{master} is numerically solved, the arbitrary phases are different for the boundary states, where we should implement gauge fixing~\cite{comsol,cui}. If we denote the real calculated eigenstate at $\mathbf{k}_{(M,j)}$ as $E'_{\mathbf{k}_{(M,j)}}(\mathbf{r})$, then,
\begin{equation}
	u'_{\mathbf{k}_{(M,j)}}(\mathbf{r}) = E'_{\mathbf{k}_{(M,j)}}(\mathbf{r})e^{-i \mathbf{k}_{(M,j)} \cdot \mathbf{r}}.
	\label{uK'}
\end{equation}
Combining \eqref{uK} with \eqref{uK'}, we have
\begin{align}
\begin{split}
u_{\mathbf{k}_{(M,j)}}(\mathbf{r})
&=
E_{\mathbf{k}_{(M,j)}}(\mathbf{r}) u'_{\mathbf{k}_{(M,j)}}(\mathbf{r}) /  E'_{\mathbf{k}_{(M,j)}}(\mathbf{r})\\
&=
E_{\mathbf{k}_{(1,j)}}(\mathbf{r}) u'_{\mathbf{k}_{(M,j)}}(\mathbf{r}) /  E'_{\mathbf{k}_{(M,j)}}(\mathbf{r}).
\end{split}
\end{align}

For degenerate bands, the Wilson loop is calculated using
\begin{equation}
	\begin{aligned}[b]
		W(\mathbf{k}_{(1,j)}) &= \operatorname{Im} \left[ \ln \prod_{m=1}^{M-1}
		S_{\mathbf{k}_{(m,j)} \rightarrow \mathbf{k}_{(m+1,j)}}
		\right], j=1,2,...,N;\\
		S_{\mathbf{k}_{\alpha} \rightarrow \mathbf{k}_{\beta}} &=\left[ \begin{matrix}
			{\left\langle u_{\mathbf{k}_{\alpha}}^1 | u_{\mathbf{k}_{\beta}}^1\right\rangle} & {\left\langle u_{\mathbf{k}_{\alpha}}^1 | u_{\mathbf{k}_{\beta}}^2\right\rangle} & \dots & {\left\langle u_{\mathbf{k}_{\alpha}}^1 | u_{\mathbf{k}_{\beta}}^n\right\rangle} \\
			{\left\langle u_{\mathbf{k}_{\alpha}}^2 | u_{\mathbf{k}_{\beta}}^1\right\rangle} & {\left\langle u_{\mathbf{k}_{\alpha}}^2 | u_{\mathbf{k}_{\beta}}^2\right\rangle}  & \dots & {\left\langle u_{\mathbf{k}_{\alpha}}^2 | u_{\mathbf{k}_{\beta}}^n\right\rangle}  \\
			\vdots & \vdots & \ddots & \vdots \\
			{\left\langle u_{\mathbf{k}_{\alpha}}^n | u_{\mathbf{k}_{\beta}}^1\right\rangle} & {\left\langle u_{\mathbf{k}_{\alpha}}^n | u_{\mathbf{k}_{\beta}}^2\right\rangle} & \dots & {\left\langle u_{\mathbf{k}_{\alpha}}^n | u_{\mathbf{k}_{\beta}}^n\right\rangle}  \\
		\end{matrix} \right]
		\label{calWL2}
	\end{aligned}
\end{equation}
where the superscript $n$ of $u^n_{\mathbf{k}}$ indicates the band index. Similarly, a gauge fixing is needed for $u_{\mathbf{k}_{(M,j)}}$.

From Fig.~\ref{cyl}(b), we can see that the first two bands are degenerate at the $K$ point and the second and third bands are degenerate at the $\Gamma$ point. Therefore, we need to compute the first three bands together by using \eqref{calWL2}. We find that a discretization of the BZ along $\bm{b_1}$ into ten plaquettes can guarantee the convergence of the results. The calculated Wilson loops corresponding to the first three bands are shown in the Fig.~\ref{WL}. For the Hermitian case, the results are symmetric about zero, and the Wannier centers of the Wilson loops are localized at the edge of the unit cell ($W = \pm \pi$), which verifies the topological nature. Same feature is observed for the PT-symmetric system with small gain and loss [middle panel of Fig.~\ref{WL}(b)]. When the gain and loss are large enough to close the band gap, the Wilson loop is disturbed from the $\Gamma$ point, as shown in the right panel in Fig.~\ref{WL}(b). For the lossy system, as in Fig.~\ref{WL}(c), the Wilson loops move downward compared with the results in the Hermitian system and the displacement has a positive correlation with the loss. Except for that, the lossy system shows similar properties as the Hermitian system, which implies that the lossy system has almost the same topological feature as the original system. Quite different from the PT-symmetric system, the PT-asymmetric system starts to show different topological feature when small gain and loss are introduced. The Wilson loops are distorted from the $\Gamma$ point, and are fully out of the original shape when gain and loss are further increased.

\section{Edge states analysis}

The interface between two PCs with different topologies supports topologically protected edge states. In Hermitian systems, the edge states have real eigenfrequencies. When gain and loss are introduced, the edge states are expected to show additional features resulting from the non-Hermiticity of the system~\cite{fan2019PT}.

\begin{figure*}[htbp]
	\centering
	\includegraphics[width=2\columnwidth]{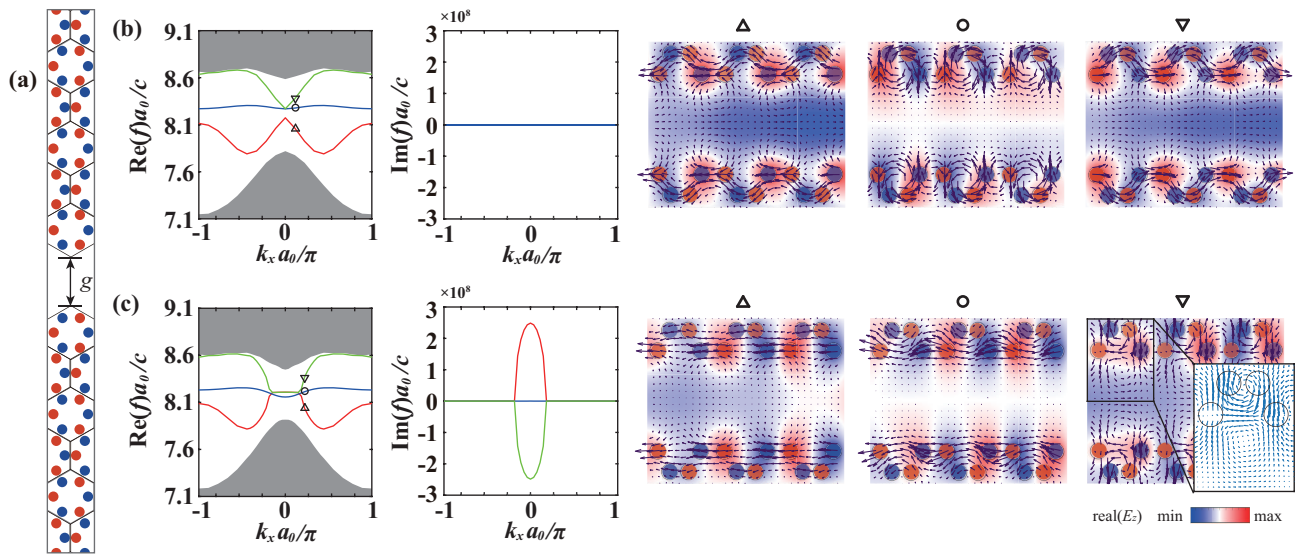}
	\caption{Topological line defect modes. (a) Supercell constructed for the topological waveguide ($a_0=2.8h$ and $g=16.4$~mm). The complex band structures and corresponding eigenstates (electric fields and time-averaged Poynting vectors) of the waveguide when (b) $\gamma_A=\gamma_B=0$ and (c) $\gamma_A=-\gamma_B=1$.}
	\label{edge_band}
\end{figure*}

We analyze a topological waveguide that is constructed by introducing an air gap in the topologically nontrivial PC (Fig.~\ref{cyl}) as shown in Fig.~\ref{edge_band}~\cite{menglinTAP}. When the system is Hermitian, three bands are identified within the band gap with real eigenfrequencies. The top and bottom bands in Fig.~\ref{edge_band}(b), have large group velocities around $k_x=0$ with states possessing even symmetry along the center of the structure, while the whole middle band has nearly zero group velocities with states possessing odd symmetry. We can see from the distributions of Poynting vectors that the energy flows concentrate at the PC-air interfaces. The net energy flows point to $-x$ and $x$ directions for the two states marked by triangles and inverted triangles, in accordance with their slopes at the band structure. Interestingly, the energy flows wind along consecutive half-orbit paths through the dielectric area reversely for the two states. Hence, the two bands represent two pseudospin channels, similar to the QSH effect. In contrast, the odd-symmetric states are just trivial defect states resulting from the symmetry of the structure.

When gain and loss increase incrementally, only the eigenfrequencies of top and bottom bands become complex conjugate pairs, starting from $k_x=0$, while the middle band keeps real [Fig.~\ref{edge_band}(c)]. The symmetries of the states keep unchanged as in the Hermitian system, but the energy flows change. Nevertheless, the orbital directions of the states of the top and bottom bands along the dielectric area are the same as in Fig.~\ref{edge_band}(b). Therefore, in the non-Hermitian system, the two bands of even-symmetric states can be classified based on the pseudospin polarizations and the band of trivial defect states keeps real.

\section{Conclusion}

In conclusion, we have studied the effects of gain and loss on the band structures of a topological PC structure and its interface with air using the generalized FD method~\cite{code}. Three cases have been considered, i.e., PT-symmetric, PT-asymmetric and lossy systems. We found that while the PT-asymmetric and lossy systems show no signature of phase transition when increasing the strength $\gamma$ of gain and loss, the PT-symmetric system shows a phase transition from the PT exact phase to PT broken phase at around $\gamma=1.3$. Moreover, the method of Wilson loop is adopted to characterize different PC systems. It has been found that the topological feature keeps for PT-symmetric and lossy systems when the band gap is kept. More importantly, the nontrivial edge states in PT-symmetric system show complex conjugate pairs of eigenfrequencies compared to its Hermitian counterparts, while the trivial defect states always show real eigenfrequencies.

\section*{Acknowledgments}

This work was supported in part by the Research Grants Council of Hong Kong GRF 17209918, AOARD FA2386-17-1-0010, NSFC 61271158,  NSFC 61975177, HKU Seed Fund 201711159228, and Thousand Talents Program for Distinguished Young Scholars of China.

\end{document}